\documentclass{webofc}
\usepackage[varg]{txfonts}  
\usepackage{graphics}
\usepackage{amsmath,amssymb}

\newcommand{\be}{\begin{equation}}
\newcommand{\ee}{\end{equation}}
\newcommand{\bea}{\begin{eqnarray}}
\newcommand{\eea}{\end{eqnarray}}

\begin{document}

\title{Light and compressed gluinos at the LHC via string theory}

\author{S.S. AbdusSalam\inst{1,2}
  \thanks{\emph{Email:} abdussalam@sbu.ac.ir}
}

\institute{Department of Physics, Shahid Beheshti University, Tehran
  19839, Islamic Republic of Iran. \and INFN, Sezione di Roma, c/o
  Department of Physics, University of Rome "La Sapienza", Italy.
}

\abstract{In this article, we show that making global fits of string
  theory model parameters to data is an interesting mechanism for
  probing, mapping and forecasting connections of the theory to real
  world physics. We considered a large volume scenario (LVS) with
  D3-branes matter fields and supersymmetry breaking. A global fit of
  the parameters to low energy data shows that the set of LVS models
  are associated with light gluinos which are quasi-degenerate with
  the neutralinos and charginos they can promptly decay into and thus
  possibly hidden to current LHC gluino search strategies.  
}

\onecolumn
\maketitle

\section{Introduction}
\label{intro}
Flux compactification with stabilised moduli sets a connection of
string theory models to real world physics \cite{hep-th/0610102,
  hep-th/0701050}. The parameters describing the various sets of
string theory models lead to different physical properties. This is a
feature associated to the so-called ``landscape'' of string theory
vacua phenomenon. The KKLT \cite{hep-th/0301240,hep-th/0503216} and large volume scenarios
\cite{hep-th/0502058,hep-th/0610129} are famous
examples with each representing sets of string theory models. Within each
of these, there are additional model variations including differences
in gauge groups, particle representations and the cosmological
constants for representing the physical world. Our target in this
article is about relating the sets of models' predictions to collider
observables and, eventually, to collider phenomenology.

The vacuum that we inhabit should be just one from the set of
landscape vacua possibilities. An interesting question goes as
follows. Is it possible to have an explicit set of string theory based
models which is in complete agreement with our real world? Here real
world can be taken to mean the standard models of particle physics and
cosmology, or even including their extensions such as with
supersymmetry breaking. For the later case, if one assumes that
supersymmetry exists in nature but broken at some energy scale
accessible to colliders, then predictions for 
supersymmetry-breaking from flux compactification models can be used
for assessing the string theory models'
parameters \footnote{Throughout this article, we assume that it is
  possible to find a compactification with exactly the MSSM matter
  content. See \cite{Braun:2005nv, Cicoli:2013mpa, Cicoli:2012vw,
    Romao:2015jrh} for instance and develoments along this
  direction.}. This is also true 
for cosmological observables that can be related to moduli 
fields. But in this article, the focus is on the particle physics
aspect.
  
Sampling string theory based model parameters for making global fit to
certain experimental data, representing the real world, could be the
best way for finding explicit models in agreement with
observations. For this purpose, here, the Bayesian techniques for
fitting models to data \cite{AbdusSalam:2008uv, AbdusSalam:2009qd}
were employed and applied to the string 
theory models. Based on the experimental results used, statistical
weights will be assigned to points for mapping and making forecasts
about the parameter space. This approach is new within string theory
phenomenological research (see \cite{1612.01569} for an overview).

For instance, consider a IIB string theory model with the dilaton and complex
structure moduli stabilised via flux superpotential $W$
\cite{hep-th/9906070, Giddings:2001yu}. Its  
vacuum expectation value (vev), $W_0$, affects the physical properties 
of the string vacuum such as the magnitude of the cosmological
constant and scale of supersymmetry breaking. A scan over the 
landscape of flux compactifications can be achieved by varying the
value of the flux superpotential $W_0$. In \cite{AbdusSalam:2007pm},
by considering a particular Calabi-Yau manifold with two moduli fields
and starting from the large volume scenario (LVS)
\cite{hep-th/0502058,hep-th/0610129} limit, with $W_0 \sim 1$, or the
KKLT limit, with $W_0 \ll 1$ \cite{hep-th/0301240,hep-th/0503216} a
set of flux-dependent AdS and dS vacua (without uplift 
terms) were obtained. Some of the minima have supersymmetry
spontaneously broken by the fluxes for matter fields on D7 branes. One
can go further by computing the full-fledged supersymmetry spectrum
via RG-running of the predicted supersymmetry-breaking terms to the
electroweak scale such as in \cite{0704.3403}. Given the supersymmetry
spectrum, other observables  
such the Higgs boson mass, dark matter relic density, electroweak
precision and B-physics observables can be computed. A non-agreement
between the predicted observables and their corresponding
experimentally measured values or limits can be used for marking the
point $W_0$ as not possible for representing our world. This way,
statistical methods for exploration and fitting the landscape
parameters can be used for mapping and physics forecasting.

In this article, we consider a particular LVS model
\cite{Aparicio:2015psl} with 
matter fields on D3-brane and supersymmetry broken via a chiral
superfield $X$ with nilpotent constraint ($X^2=0$)
\cite{Kallosh:2014wsa, Bergshoeff:2015jxa, Kallosh:2015nia}. 
 In section \ref{sec:lvs} we
briefly introduce the LVS and the supersymmetry-breaking terms derived
in \cite{Aparicio:2015psl} for setting the context of the article and
defining an LVS subset of the minimal supersymmetric standard model
(MSSM), with R-parity conserved, called LMSSM-6 as the
phenomenological frame for our analysis. The Bayesian method for
fitting the LMSSM-6 to low-energy physics is presented in
section~\ref{lmssm6fit}. The results of the global fit indicate an
abundance of light gluinos, neutralinos and
charginos together with ${\cal O}(10) TeV$ squarks. The
electroweak-inos are within one to few top-quark heavy. The mass
difference between the gluinos, charginos or heavy neutralinos and the
lightest supersymmetric particle (LSP) turns out to be small thereby
making a lower sub-TeV quasi-degenerate spectrum. As such the gluinos and
electroweak-inos are possibly surviving current LHC limits. Detailed
reinterpretation or recasting analyses should ultimately lead to a
robust conclusion about the status of such spectrum at the LHC. In
section \ref{sec:conc} we round up the the article with conclusions
and outlook.  

\section{Large volume scenario (LVS) SUSY-breaking, LMSSM-6}
\label{sec:lvs}
The shape and volume of string theory internal space
represent massless scalar fields from a $4$-dimensional theory point
of view. These moduli fields must be stabilised since no such extra
scalar fields are observed in nature. There are many moduli
stabilisation work in the literature. Here we concentrate on a LVS scenario where all the
moduli fields are fixed at an exponentially large internal volume. 
The LVS supergravity effective theory couples
to the particle content of the MSSM. In the effective theory, the K\"ahler and
superpotentials are generated from the superstring theory. This forms
the hidden sector which then couples via gravity mediation
to the MSSM sector. 4d $N=1$ supergravity is
specified up to two derivatives by the K\"ahler 
potential $K$, superpotential $W$ and gauge kinetic function
$f_a$. With these, the scalar potential is given by 
\bea
V &=& e^K \left[ G^{i\bar{j}} D_i W \overline{D_j W} -
  3|W|^2\right], \\
D_i W &=& \partial_i W + (\partial_i K)W,\\
G^{i\bar{j}} &=& (\partial_i \partial_{\bar{j}} K)^{-1}.
\eea
Here $i, j$ run over the two  K\"ahler moduli $T_s$ and $T_b$ whose
real parts  $\tau_b$ and $\tau_s$ determine the internal space volume, 
$\mathcal{V} = \tau_b^{3/2}-\tau_s^{3/2}$. 
At the minimum of the potential the moduli fields acquire vevs and
provide non vanishing auxiliary field that spontaneously breaks
supersymmetry and generate the breaking terms in the visible
sector. Other dynamics such as the presence of D3-brane within the
superstring construction can also lead to the breaking of
supersymmetry.

The LVS supersymmetry-breaking terms induced by the
presence of nilpotent superfield $X$ \cite{Aparicio:2015psl} is
presented as follows. The K\"ahler potential and superpotentials for
$X$ can be taken as
\be
K=K_0  + K_1X + \bar{K}_1\bar{X}\ + K_2X\bar{X} , \qquad W=\rho X +
W_0. 
\ee
where $K_0, K_1, K_2, \rho, W_0$ are coefficients and can be functions
of other low-energy fields.
Assuming the dilaton and complex structure moduli have been fixed and
integrated out at high scale, the LVS K\"ahler potential becomes 
\begin{equation}
  K = -2\log \left(\mathcal{V}- \hat\xi \right)  
  + \tilde{K}_i\ \phi \bar{\phi} + \tilde{Z}_i\ X \bar{X} +
  \tilde{H}_i\ \phi \bar{\phi} \ X \bar{X} + ... \,.
  \label{Klvs}
\end{equation}
Here $\hat\xi = \frac{s^{3/2}}\xi{2}$, $\xi$ is a Calabi-Yau manifold
related constant of order one \cite{Becker:2002nn}. $s=1/g_s$ is the
real-part of the axion-dilaton field. $\phi$ represents matter,
$\tilde{K}_i$ and $\tilde{Z}_i$ are respectively the D3-branes matter
and the nilpotent goldstino metrics. $\tilde{H}_i $ represents the
nilpotent goldstino and matter fields quartic interaction. These are
parameterised for the $\alpha^{'}$-corrected potential as 
\begin{equation}
  \tilde{K}_i = \frac{\alpha_0}{\mathcal{V}^{2/3}}\left(1-\alpha_1 \frac{\xi s^{3/2}}{\mathcal{V}} \right),
  \label{mattermetric2}
  \quad
  \tilde{Z}_i = \frac{\beta_0}{\mathcal{V}^{2/3}}\left(1-\beta_1 \frac{\xi s^{3/2}}{\mathcal{V}} \right),
  \quad
  \tilde{H}_i = \frac{\gamma_0}{\mathcal{V}^{4/3}}\left(1-\gamma_1 \frac{\xi s^{3/2}}{\mathcal{V}} \right).
\end{equation}
The
LVS superpotential is given by  
\begin{equation}
  W = W_0 + \rho X  + A e^{-a_s T_s}
\end{equation}
where $A$, and $a_s$ are gaugino condensation parameters. 
The K\"ahler and superpotentials can then be used for computing
the soft supersymmetry-breaking terms via a standard method as done in
\cite{Aparicio:2015psl}. The scalar $m_0$, gaugino $M_{1/2}$, and trilinear coupling $A_0$ soft 
supersymmetry-breaking terms for visible sector fields on 
D3-brane are 
\begin{equation} \label{m0term}
  m_0^2 = \frac{5}{4} \frac{s^{3/2} \xi}{\mathcal V} \, (3
  \alpha_1 -1)\, m_{3/2}^2 + \frac{9}{8}\frac{s^{3/2} \,
    \xi}{\mathcal{V}} \, \frac{1}{5a_s \tau_s} \, \left(1-\frac{3
    \gamma_0}{\alpha_0 \beta_0} \right) m_{3/2}^2 , 
\end{equation}
\begin{equation}\label{Mterm}
  M_{1/2} = sign(W_0) \, \frac{3}{4}\frac{s^{3/2}\xi}{\mathcal V} \left[ 3 -
    2\omega_s \right]\,m_{3/2}, \textrm{ and }
\end{equation}
\begin{equation} \label{a0term}
  A_0 = - (1- y) M_{1/2} \;, y = s\partial_s \log Y_{ijk}^{(0)}.
\end{equation}
Here $m_{3/2} = e^{K/2} |W|$ is the gravitino mass.
$\omega_s \lesssim 1$ parametrise corrections such that
$D_sW \sim \frac{e^{-a\,\tau}}{2s}\omega_s$ used in computing the soft
supersymmetry-breaking terms. 

Given a set of LMSSM-6 parameters, the supersymmetry-breaking terms
Eq.(\ref{m0term}) to Eq.(\ref{a0term}) can be computed. These are then set as
the boundary conditions for renormalisation group (RG) running from
the supersymmetry-breaking 
scale, $m_{3/2}$, to the weak-scale. For doing this we chose a particular
specialisation of gravity-mediated supersymmetry-breaking with
a non-universal scalar mass terms for the MSSM Higgs doublets such
that $m_{H_1} = m_{H_2} = 0$ at symmetry-breaking scale. Other
possibilities include varying the non-universal Higgs doublet
mass terms and the minimal supergravity but are not considered
here. The set of LVS parameters,
$\{ \, x, \, \alpha_1, \, y, \, \omega_s, \, \tan \beta, \,
\log_{10} m_{3/2} \}$,  in the soft supersymmetry-breaking terms above
together the RG to the weak scale with MSSM sparticle content is
referred to as the LVS MSSM with six parameters (LMSSM-6). $\tan \beta$
is the ratio of the Higgs doublet vevs. For making a
global fit of the LMSSM-6 parameters to data we add 5 Standard Model
nuisance parameters that are for some of the precision observables
used. As such we will be exploring a total of 11 parameters
\begin{equation} \label{lmssm11}
  \underline{\theta} \equiv \{ \, x, \, \alpha_1, \, y, \, \omega_s, \,
  \tan \beta, \, \log_{10} m_{3/2}, m_Z, \, m_t, \, m_b, \, \alpha_{em},\, \alpha_s \}. 
\end{equation}
The Standard Model nuisance parameters are the
Z-boson mass  $m_Z = 91.1876 \pm 0.0021$ GeV, the top quark mass
$m_t = 172.6 \pm 1.4$ GeV, the bottom quark mass
$m_b = 4.2 \pm 0.07$ GeV, the electromagnetic coupling constant
$\alpha_{em}^{-1} = 127.918 \pm 0.018$, and the strong interactions
coupling constant $\alpha_s = 0.1172 \pm 0.002$. These were all set to
vary in a Gaussian manner with central values and deviations according
to the experimental results. The LVS parameters are were allowed to vary as: 
$x \equiv \frac{3 \gamma_0}{\alpha_0 \beta_0} \in [0.01, 100.0]$,
$\alpha_1 \in [0.01, 100.0]$,
$y \in [-1000, 1000]$,
$\omega_s \in [0.01, 100.0]$,
$\tan \beta \in [2, 60]$,
$log_{10} m_{3/2} \in [3, 19]$. The other LVS parameters which were
not varied but enter the soft terms are $\xi=1.0, W_0=\pm 1.0,s=25.0$, 
and ${\mathcal V}=1.0 \times 10^{9}$ units. In the next section we describe
the procedure for fitting the parameters to low energy data. 

\section{The LMSSM-6 fit to low energy physics}
\label{lmssm6fit}
We consider a flat (Gaussian) prior probability density distribution,
$p(\underline \theta )$, for the LMSSM-6 (Standard Model) parameters
in Eq.(\ref{lmssm11}) for exploring the space using the
{\sc MultiNest} \cite{Feroz:2007kg,Feroz:2008xx} package which
implements Nested Sampling algorithm~\cite{Skilling}. During the model
exploration, 
at each parameter space point the following packages were used via the
SLHA1~\cite{Skands:2003cj} interface: the
supersymmetry spectrum generator {\sc
  SOFTSUSY}~\cite{Allanach:2001kg} for computing the 
sparticle masses, mixing angles and couplings, {\sc
  micrOMEGAs}~\cite{Belanger:2008sj} for 
computing the neutralino cold dark matter relic density and the
anomalous magnetic moment of the muon $\delta a_\mu$, {\sc
  SuperIso}~\cite{Mahmoudi:2007vz} for predicting the branching
ratios $BR(B_s \rightarrow \mu^+ \mu^-)$, $BR(B \rightarrow s \gamma)$
and the isospin asymmetry, $\Delta_{0-}$ in the decay
$B \rightarrow K^* \gamma$, and {\sc
  susyPOPE}~\cite{Heinemeyer:2006px,Heinemeyer:2007bw} 
for computing precision observables that include the $W$-boson mass
$m_W$, the effective leptonic mixing angle variable $\sin^2
\theta^{lep}_{eff}$, the total $Z$-boson decay width, $\Gamma_Z$, and
the other electroweak observables whose experimentally
determined central values and associated errors are summarised in
Table~\ref{tab:obs}. 
The experimental central values  
($\mu_i$) and errors ($\sigma_i$) for these make the set of the predictable observables, 
$\underline O$:  
\bea \label{observs}
\underline O &\equiv &\{ m_h, \; m_W,\; \Gamma_Z,\; \sin^2\,
\theta^{lep}_{eff},\; R_l^0,\; R_{b,c}^0,\;
A_{FB}^{b,c},\;  A^l = A^e,\; A^{b,c}, \\ \nonumber
& &BR(B \rightarrow X_s \, \gamma),\; BR(B_s \rightarrow \mu^+ \,
\mu^-),\; \Delta M_{B_s},\;  R_{BR(B_u \rightarrow \tau
  \nu)},\\ \nonumber
& &\Omega_{CDM}h^2, \; Br(B_d \rightarrow \mu^+ \mu^-), \; \Delta
M_{B_d} \}.
\eea
\begin{table}
  \begin{center}
  \caption{Summary for the central values and errors for the electroweak
    precision, B-physics and cold dark matter relic density
    constraints.}
  \label{tab:obs}
  \begin{tabular}{|ll||ll|}
    \hline\noalign{\smallskip}
    Observable & Constraint & Observable & Constraint  \\
    \hline
    $m_W$ [GeV]& $80.399 \pm  0.023$ \cite{verzo}&$A^l = A^e$& $0.1513 \pm
    0.0021$ \cite{:2005ema} \\
    $\Gamma_Z$ [GeV]& $2.4952 \pm 0.0023$ \cite{:2005ema}&$A^b$ & $0.923
    \pm 0.020$ \cite{:2005ema}\\
    $\sin^2\, \theta_{eff}^{lep}$  & $0.2324 \pm 0.0012$ \cite{:2005ema}&$A^c$ & $0.670 \pm 0.027$ \cite{:2005ema}\\
    $R_l^0$ & $20.767 \pm 0.025$ \cite{:2005ema} &$Br(B_s \rightarrow
    \mu^+ \mu^-)$ & $3.2^{+1.5}_{-1.2} \times 10^{-9}$
    \cite{Aaij:2012nna}\\
    $R_b^0$ & $0.21629 \pm 0.00066$ \cite{:2005ema}&$\Delta M_{B_s}$ &
    $17.77 \pm 0.12$  ps$^{-1}$ \cite{Abulencia:2006ze}\\
    $R_c^0$ & $0.1721 \pm 0.0030$ \cite{:2005ema}&$R_{Br(B_u \rightarrow \tau \nu)}$&
    $1.49 \pm 0.3091$ \cite{Aubert:2004kz}\\
    $A_{\textrm{FB}}^b$ & $0.0992 \pm 0.0016$ \cite{:2005ema}& $\Delta
    M_{B_d}$ & $0.507 \pm 0.005$ ps$^{-1}$\cite{Barberio:2008fa} \\
    $A_{\textrm{FB}}^c$ & $0.0707 \pm 0.0035$ \cite{:2005ema}&$\Omega_{CDM} h^2$ & $0.11
    \pm 0.02 $ \cite{0803.0547}\\
    $m_h$ [GeV] & $125.6 \pm 3.0$ \cite{1207.7214, 1207.7235} &
    $Br(B\rightarrow X_s \gamma)$ & $(3.52 \pm 0.25) \times 10^{-4}$
    \cite{Barberio:2007cr}\\
    \hline
  \end{tabular}
\end{center}
\end{table}

The compatibility of the LMSSM-6 parameter-space points with the data is
quantified by the likelihood,
$p(\underline d|\underline \theta)$. Assuming the observables are
independent, the combined likelihood can calculated as 
\be p(\underline d|\underline \theta) = L(x) \,
\prod_i \, \frac{ \exp\left[- (O_i - \mu_i)^2/2
    \sigma_i^2\right]}{\sqrt{2\pi \sigma_i^2}}
\ee where the index $i$ runs over the list of observables
$\underline O$. Here, $x$ represents the predicted value
of neutralino cold dark matter (CDM) relic density and 
\be \label{olik}
L(x) =
\begin{cases}
  1/(y + \sqrt{\pi s^2/2}) &  \textrm{ if } x < y \\
  \exp\left[-(x-y)^2/2s^2\right]/(y + \sqrt{\pi s^2/2}) &
  \textrm{ if } x \geq y
\end{cases}
\ee
where $y = 0.11$ is the CDM relic density central value and $s=0.02$ 
the corresponding inflated (to allow for theoretical uncertainties)
error. 

\begin{figure}
  \begin{center}
    \resizebox{0.75\textwidth}{!}{\includegraphics{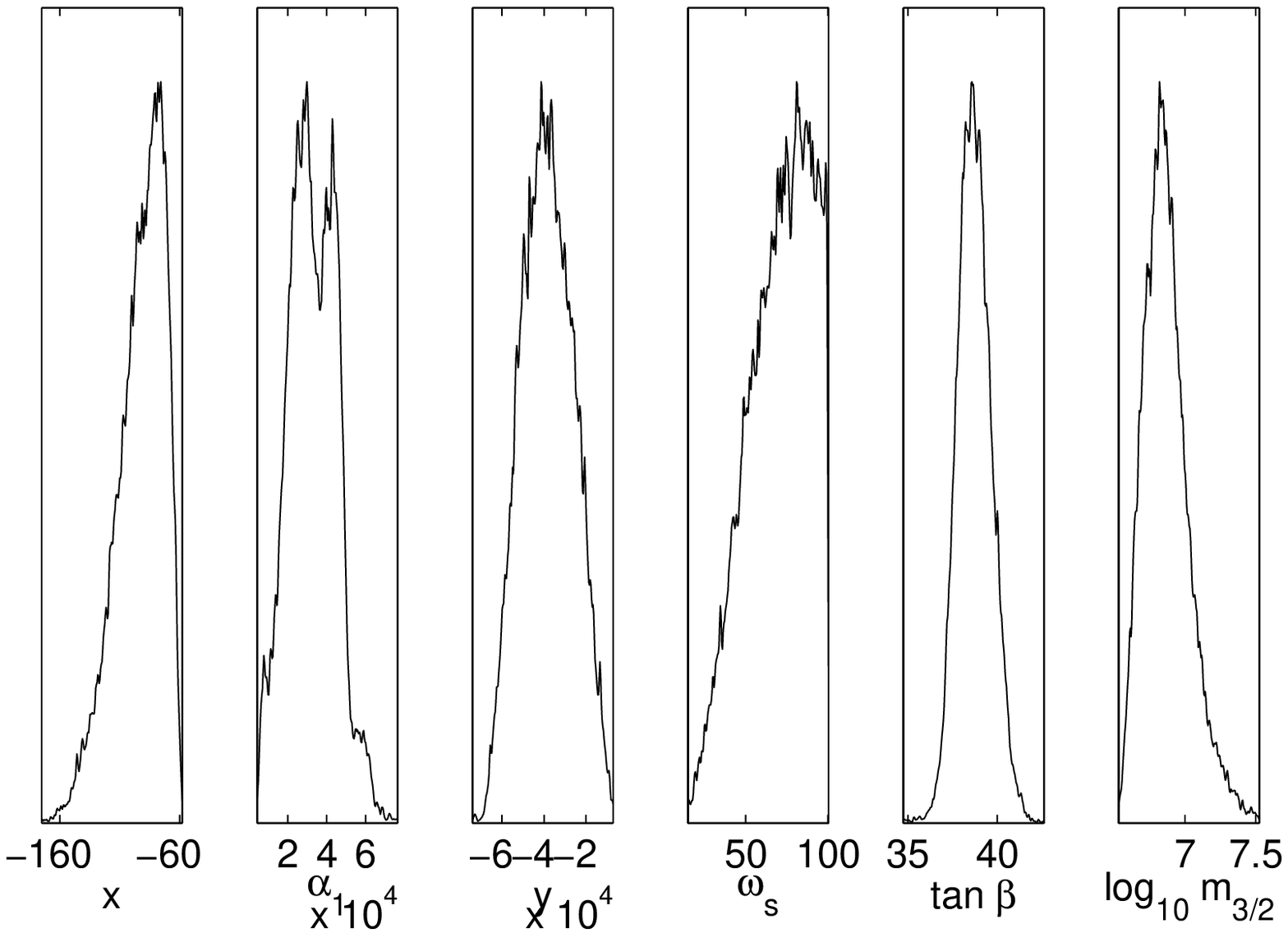}} \\
    \resizebox{0.75\textwidth}{!}{\includegraphics{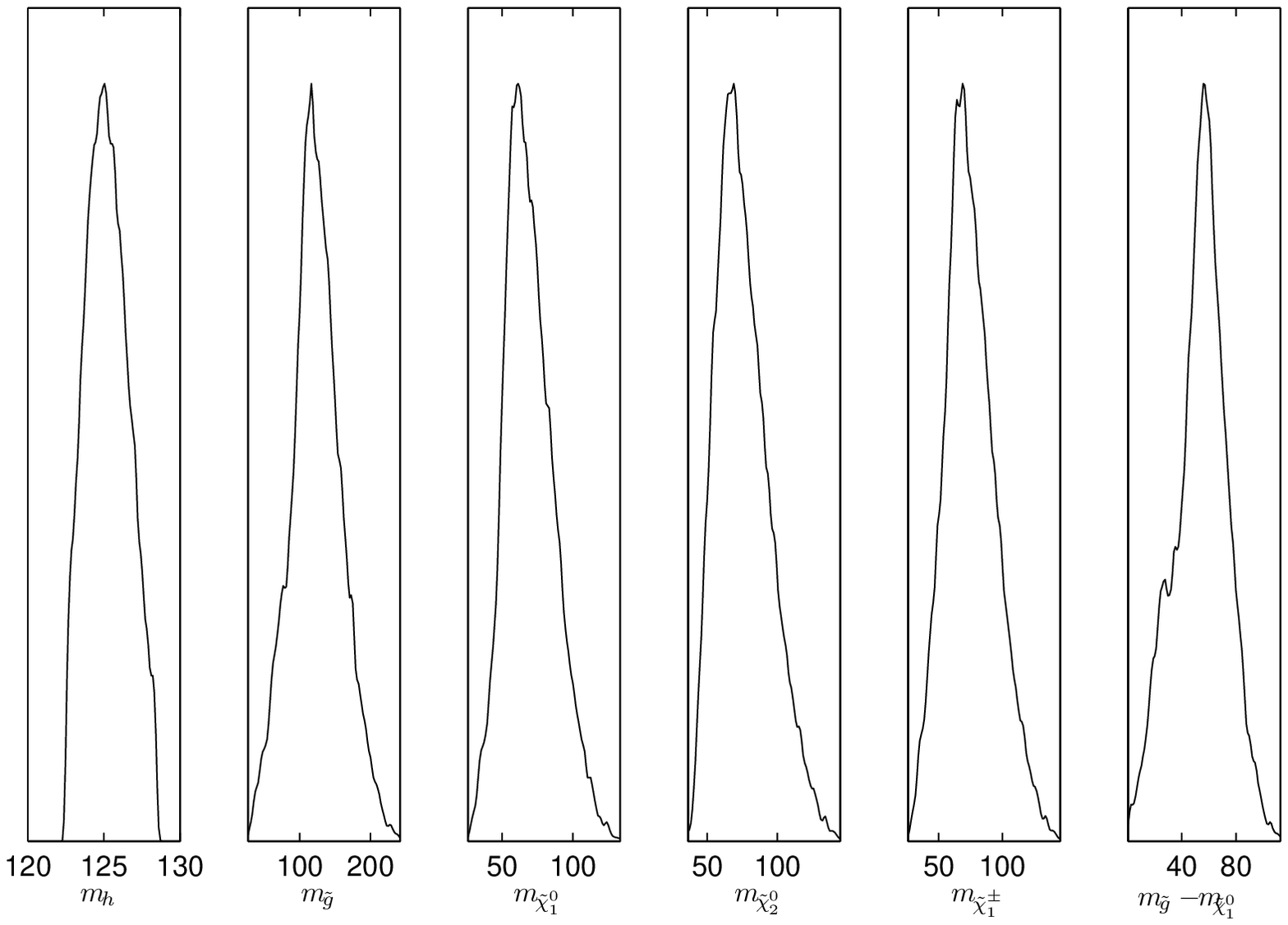}}
    \caption{First row: 1-dimensional posterior probability
      distributions for the SUSY-breaking parameters and (second row)
      for the Higgs boson and electroweak-inos mass in GeV units. Corresponding 
      squark masses (not shown here) are all of order 10s of TeV. The
      vertical axis is a measure of the relative posterior
      probabilities between the x-axis points. 
    }
    \label{fig:i}
  \end{center}
\end{figure}

\begin{figure}
  \begin{center}
    \resizebox{0.65\textwidth}{!}{\includegraphics{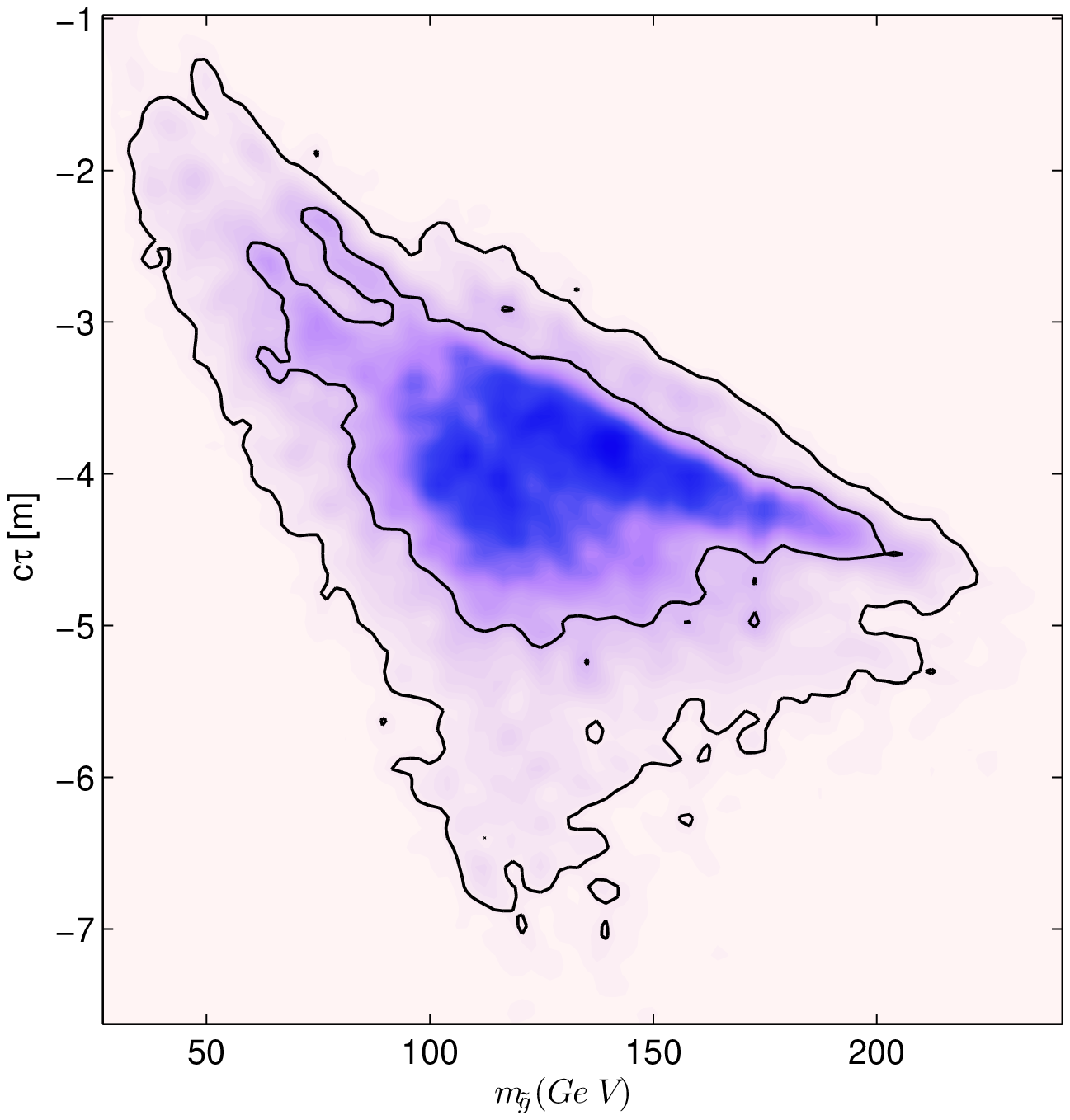}}
    \caption{Marginalised 2-dimensional posterior distribution for the
      gluino decay length versus the gluino mass. The contour lines
      show the 68\% and 95\% Bayesian credibility regions.}
    \label{fig:lifetime1}
  \end{center}
\end{figure}

The outcome of the Bayesian fit of the LMSSM-6 to data is the
posterior probability density
\be \label{bayes}
p(\underline \theta |\underline d) = \frac{p(\underline d|\underline
  \theta) \, \times \, p(\underline \theta)}{p(\underline d)}
\ee
shows that the sfermions are order tens of TeV heavy
while the gluino and electroweak-inos are light, in few hundreds of
GeV range. 
The posterior probability distributions for the LMSSM-6
supersymmetry-breaking parameters, the Higgs boson mass for reference
purpose, the electroweak-inos and the gluino-neutralino mass
difference are shown in Figure~\ref{fig:i}. 
The heavy squarks can potentially make the gluinos long-lived or to turn
into R-hadrons before decaying. In Fig.~\ref{fig:lifetime1}, the
2-dimensional posterior distribution for gluino decay length estimate
versus its mass is given. The gluino decay life-time for squark at
order 10 TeV is estimated as \cite{Dawson:1983fw,Hewett:2004nw}
\begin{equation}
  \tau \approx 8 \, \left( \frac{\tilde m}{10^9 \, \textrm{ GeV}
  }\right)^4 \, \left( \frac{ 1 \, \textrm{ TeV} }{m_{ \tilde g}
  }\right)^5 \,\, s
\end{equation}
It turns-out that the gluino and electroweak-inos are compressed
relative to the LSP mass. For this reason, one cannot automatically
rule out LMSSM-6 light gluinos. 

The direct search for gluinos at the LHC
puts the lower limit on gluino mass well into the TeV towards
multi-TeV region (see for instance refs
\cite{1507.05525,SUS-16-014,SUS-16-015}.)
The limits come from search channels with multijets plus
zero leptons and can probe gluinos decaying into gluons and
neutralinos. The same is the case for the gluino decay to two quarks
and a neutralino; and for the topologies with initial state
radiations. However it is not clear what the status will be for
the light gluinos which are quasi-degenerate with the neutralinos or 
charginos they can decay to. Gluino production cross-section at the
LHC is much higher than squarks' of the same mass. So in principle as
long as the gluino decays to gluon and neutralino or to a quark pair
and neutralino then then, for instance, the CMS limits from
searches with multi-jet final states \cite{SUS-16-014,SUS-16-015}
should apply for the LMSSM-6 gluinos. But, that and the application of
other relevant LHC limits necessarily require rigorous and
dedicated reinterpretation analysis of the results.

In \cite{1701.07664} various experimental search strategies
for finding gluinos quasi-degenerate with neutralinos have been
addressed. These include searches for displaced vertices and disappearing
tracks. The analysis was however based on a fixed gluino mass at
1.5~TeV. It will be interesting to see what the outcome of a similar
analysis will be for light gluinos such as for the LMSSM-6 shown in
this article. We again emphasis that a decisive conclusion requires
detailed reinterpretation of the experimental results within the 
LMSSM-6 context.

\section{Conclusion and outlook} 
\label{sec:conc}
Models from string theory compactification with moduli stabilised by 
fluxes can predict supersymmetry-breaking at the TeV scale. The
generation 
of supersymmetry spectrum allows the connection of the string theory or
landscape parameters to experimental observables for constraining the
the theory. In this article, we have applied Bayesian global fit
technique for string theory inspired phenomenology. We introduce LMSSM-6 for
representing the set of supersymmetry-breaking parameters predicted
from a large volume scenario with matter fields on D3-branes. The field content were
considered to be that of the minimal supersymmetric standard model with
R-parity conserved and neutralino lightest supersymmetric
particle as dark matter candidate. This way, the LMSSM-6 parameters
which originate from the string theory setting can be constrained via
the low energy properties of the associated sparticle spectrum. This
feature was used for forecasting and mapping the string theory based
parameter space. 

The low energy constrains used are the Higgs boson mass, dark matter relic
density, electroweak precision and B-physics experimentally measured
observables. The global fit of the LMSSM-6 parameters to these showed that
the bulk of the posterior distribution yielded
heavy squarks at order 10 TeV. These come together with 
light (much less than or of the order of the 
top-quark mass) but promptly decaying gluinos. 
Such gluinos would have been already ruled out even before the LHC if
not for being  quasi-degenerate with neutralinos and charginos. A
dedicated reinterpretation of the experimental results that probe
gluino productions at the LHC such as in
\cite{1507.05525,SUS-16-014,SUS-16-015} is needed for determining the
status of the gluinos. This is an
interesting work to be done but is beyond the scope of this article. 
Should the LMSSM-6 gluinos be ruled out, then the corresponding class
of large volume scenario models cannot represent our real world. 

As an outlook, the following are interesting questions or work to be
done based on the concepts addressed in this article. 
\begin{itemize}
\item The methodology presented can be applied to other string theory
  phenomenology frames such as the famous KKLT
  \cite{hep-th/0301240,hep-th/0503216} scenarios in comparison to the
  large volume ones.
\item The posterior sample from the global fit of parameters to data
  can be used in recasting collider results for 
  establishing the status of the considered models (such as the
  LMSSM-6 presented here).
\item There is an interesting complementarity between the
  supersymmetry spectrum such as from the LMSSM-6 global fit to data
  presented in this article, and similar spectra that could be
  constructed via the simplified models approach to supersymmetry
  phenomenology. With the global fit approach, the resulting spectra
  are guaranteed to be in agreement with the experimentally measured
  valued of the observables used for the fitting procedure. This is
  not necessarily the case for simplified model spectra. It is
  interesting to explore further and contrast this and similar
  characteristics.   
\end{itemize}

\section*{Acknowledgment} Thanks to M.Pierini and F.Quevedo for
discussions; and ICTP for hospitality during an early stage of the
research presented here. An early stage of this research was supported
by an ERC fund under the European Union's Seventh Framework Programme
(FP/2007-2013)/ERC Grant Agreement no. 279972 NPFlavour.

\end{document}